\documentclass[12pt,preprint]{aastex}
\usepackage{amsmath}
\usepackage{graphicx}

\usepackage{enumitem}
\usepackage{color}
\usepackage{url}                         % For breaking URLs easily trough lines
\usepackage[normalem]{ulem}
\usepackage{sidecap}

\usepackage{hhline}
\usepackage{multirow}

\slugcomment{\textcolor{white}{aaa}}

\shorttitle{CoMP Tomography}
\shortauthors{Kramar et al.}

\begin{document}


%\title{Deflection of the streamer belt toward radially directed structure}
\title{Direct Observation of Coronal Magnetic Fields by Vector Tomography of the Coronal Emission Line Polarizations}


\author{M. Kramar\altaffilmark{1}, H. Lin\altaffilmark{2}, S. Tomczyk\altaffilmark{3}}
\affil{\altaffilmark{1}Physics Department, The Catholic University of America, 
620 Michigan Ave NE, Washington, DC 20064}
%\email{kramar@cua.edu}

\affil{\altaffilmark{2}
Institute for Astronomy, University of Hawaii at Manoa, 34 Ohia Ku Street, Pukalani, Maui, HI 96768}
%\email{lin@ifa.hawaii.edu}

%\author{S. Tomczyk\altaffilmark{}}
\affil{\altaffilmark{3}High Altitude Observatory, 3080 Center Green Drive, Boulder, CO 80301}
%\email{tomczyk@ucar.edu}

%\and

\altaffiltext{1}{e-mail: kramar@cua.edu}

%\altaffiltext{2}
%{Institute for Astronomy, University of Hawaii at Manoa, 34 Ohia Ku Street, Pukalani, Maui, HI 96768}

%\altaffiltext{3}
%{High Altitude Observatory, 3080 Center Green Drive, Boulder, CO 80301}


%\altaffiltext{1}{Physics Department, The Catholic University of America, 
% 620 Michigan Ave NE, Washington, DC 20064.}
%\altaffiltext{2}{NASA-GSFC, Code 671, Greenbelt, MD 20771.}
%\altaffiltext{3}{High Altitude Observatory, Boulder, CO}
%\altaffiltext{4}{nstitute of Astronomy, University of Hawaii}


\begin{abstract}
This article presents the first direct ``observation'' of the global-scale, 
3D coronal magnetic fields of Carrington Rotation (CR) Cycle 2112 using vector tomographic inversion techniques. 
The Vector tomographic inversion uses observational measurements of the Fe~{\sc{xiii}} 10747 \AA\ Hanle effect polarization signals by the Coronal Multichannel Polarimeter (CoMP) and coronal density and temperature structures derived from scalar tomographic inversion of STEREO/EUVI coronal emission lines (CELs) intensity images as inputs to derive a coronal magnetic field model that best reproduces the observed polarization signals. 
While independent verifications of the vector tomography results cannot be performed, 
we compared the tomography inverted coronal magnetic fields with those constructed by  magnetohydrodynamic (MHD) simulations based on observed photospheric magnetic fields of CR 2112 and 2113. We found that the MHD model for CR 2112 is qualitatively consistent with the tomography inverted result for most of the reconstruction domain except for several regions. Particularly, for one of the most noticeable regions, 
we found that the MHD simulation for CR 2113 predicted a model that more closely resembles 
the vector tomography inverted magnetic fields. 
In another case, our tomographic reconstruction predicted an open magnetic field 
at a region where a coronal hole can be seen directly from a STEREO-B/EUVI image. 
We discuss the utilities and limitations of the tomographic inversion technique, 
and present ideas for future developments.
\end{abstract}



\section{Introduction}

Routine direct measurements of the coronal magnetic fields are indispensable for our eventual understanding of the nature of solar coronal phenomena at all spatial and temporal scales. Important progresses were made in the last two decades in the development of the theories \citep{Casini_1999, Lin_2000, Casini_Lin_2002}, instrumentation and observing techniques \citep{lin_et_al_2000, Lin_2004} to directly measure the polarization of coronal emission lines (CELs) that bear the information of the coronal magnetic fields. A major mile stone was reached with the deployment of the Coronal Multichannel Polarimeter \citep[CoMP,][]{Tomczyk_2007,Tomczyk_2008}. 
CoMP provides synoptic linear polarization measurements of the Fe~{\sc{xiii}} 10747 \AA\ emissions which directly maps the orientation of the magnetic fields of the solar corona. However, due to the low density and opacity of the solar atmosphere the observed polarization signals are the total signals summed along the line-of-sight (LOS) of the measurements with non-uniform temperature, density and magnetic field structures. 
Therefore, except in a few special cases, direct inference of the 3D coronal magnetic field structure from polarization data is in general not possible. 

Parallel to the development of the observing capabilities, a vector tomography inversion method to infer the coronal magnetic fields based on CEL circular polarization data was developed by \cite{Kramar_2006}. More recently a second vector tomography inversion method that combines coronal linear polarization data and information of the temperature and density structure of the corona to reconstruct the 3D coronal magnetic fields was developed by \cite{Kramar_2013}. 
We applied this technique to CoMP data obtained during Carrington Rotation (CR) cycle 2112 to derive the global 3D coronal magnetic fields up to $1.3 R_\odot$. However, as it was pointed out in \cite{Kramar_2013}, while polarizations of the CELs respond directly to the local magnetic fields, their visibility, i.e., the strength of the emission lines, strongly depends on local thermodynamic properties of the plasma.Therefore, 3D temperature and density structures of the corona are also needed as inputs for the vector tomographic inversion program for the calculation of the strength of the coronal emission lines. In this work, the 3D temperature and density structure of the corona were derived by scalar tomographic inversion of the ultraviolet coronal images obtained by {\it Extreme Ultraviolet Imager} (EUVI)  on board of the {\it Solar Terrestrial Relations Observatory} (STEREO) spacecraft. In the following sections, we will first describe the scalar inversion we used to derive the temperature and density distribution of the corona from STEREO/EUVI data. Then we will present results from the vector tomographic inversion of the 3D magnetic fields of CR 2112. Limitations of this methods, as well as future improvement of the technique will be discussed. 

\section{Reconstruction of Density and Temperature}
\label{Sec_Density_Temp}

%\textcolor{blue}{
The process of 
the 3D reconstruction of the coronal electron density and temperature based on STEREO/EUVI data consists of two steps. 
First, the 3D reconstruction of EUVI emissivities. 
Second, finding the density and temperature values which satisfies the previously reconstructed emissivities for every voxel in the 3D reconstruction domain. 
While first step is the same as in the Differential Emission Measure Tomography (DEMT) method described by \cite{Frazin_2005ApJ_DEMT, Frazin_2009ApJ_DEMT, Vasquez_Frazin_2010ApJ}, 
our implementation of second step is different. 
The main differences in second step are that, in our method, 
we assumed the plasma is isothermal within each volume element (voxel) and 
different implementation of an inversion procedure. 

Briefly, the measured coronal emission $I_\lambda$ in any of the EUVI channels can be represented by the LOS integral 
\begin{equation}
I_\lambda=
k \cdot \int \limits_\textrm{LOS} \varepsilon_{\lambda}(\vec{r}) d \ell ,
\label{EUVI_Emiss_LOS}
\end{equation}
where $\varepsilon_{\lambda}(\vec{r})$ is the emissivity at the position $\vec{r}$ in the corona for the selected channel $\lambda$, i.e., light intensity (in photons per second) emitted per unit volume, per unit solid angle. The coefficient $k$ accounts for pixel size, aperture, distance to the Sun.  The inversion of \ref{EUVI_Emiss_LOS}  for  $\varepsilon_{171}(\vec{r})$, $\varepsilon_{195}(\vec{r})$, and $\varepsilon_{284}(\vec{r})$ is performed by tomography method described in \cite{Kramar_2014}. 

For an locally isothermal plasma, $\varepsilon_{\lambda}(\vec{r})$ can be expressed by \citep{Phillips_2008} 
\begin{equation}
\varepsilon_\lambda = G(T,N_e) N_e^2, 
\label{EmissFun}
\end{equation}
where $T$ is the temperature, $N_e$ is electron density (and a function of $T$), and $G(T,N_e)$ is the contribution function. For our inversion $G(T,N_e)$ was calculated with the CHIANTI code \citep{Dere_2009chianti}, and the Fe ions abundances were derived from results of \cite{Arnaud_Raymond_1992}. For every grid cell, we derive three (one for each EUVI channel) $N_e(T)$ curves from (\ref{EmissFun}) based on the emissivities $\varepsilon_\lambda(\vec{r})$ derived by the tomography from equation (\ref{EUVI_Emiss_LOS}). With the local isothermal approximation, the three $N_e(T)$ curves will converge at a point indicating the solution for $T$ and $N_e$ for a given grid cell. 

We applied this inversion method to the STEREO/EUVI \citep{Wuelser_2004, Howard_2008} data of CR 2112. STEREO/EUVI observes the coronal up to approximately 1.7 $R_{\odot}$ in four spectral channels (171, 195, 284, and 304 \AA). We used the off-limb data of images from the 171, 195, and 284 channels for the tomographic inversion of the temperature and density. Similarly to \cite{Vasquez_Frazin_2010ApJ}, three average images per channel per day over half a solar rotation during CR2112 were used for the reconstruction. The average images were obtained by averaging three off-limb images taken with a two hours interval. The resolution of the EUVI images were also reduced to 512 x 512 pixels. 
The upper boundary of the reconstruction domain is limited to $1.5 R_{\odot}$ due to the lower signal-to-noise ratio of data higher in the corona. 
%\textcolor{blue}{
The lower boundary is set to $1.05 R_\odot$ %because the corona below might not be optically thin near the limb 
in order to avoid optical depth effects near the limb \citep{Frazin_2009ApJ_DEMT,Vasquez_Frazin_2010ApJ}. %}
Figure \ref{Fig_NeTe_Rec} shows a spherical cross-section of the 3D temperature and electron density for CR 2112 at $1.17 R_{\odot}$ reconstructed with this method. Additionally, the density map at $2.0 R_{\odot}$ obtained by tomographic inversion of the STEREO/COR1 data \citep{Kramar_2009} are shown. 
This map will be used in later sections to verify the magnetic field of the corona. 

\section{Reconstruction of Magnetic Fields}

The CoMP instrument measures the intensity and the linear and circular polarizations (Stokes I,Q,U,V) of the Fe XIII lines at 10747 and 10798 \AA\ of the solar corona from $\sim 1.03$ to $\sim 1.4 R_\odot$. In the present research, we use data only from the 10747 \AA\  forbidden line. Averaged daily CoMP Stokes $Q$ and $U$ images from July 13 to July 26, 2011, 
corresponded to CR 2112, are used for this study. The reconstruction of the coronal magnetic fields follows the method described in \cite{Kramar_2013}, using the 3D $T$ and $N_e$ derived from tomographic inversion of STEREO/EUVI data as fixed input parameters. Although $T$ and $N_e$ were derived out to $1.5 R_{\odot}$, examination of the results indicated that the inversion above $1.3 R_{\odot}$ may not be very reliable. Therefore the outer boundary of the reconstruction domain for the magnetic fields were limited to $1.3 R_{\odot}$ for this work. 

\begin{figure}[h]
\includegraphics*[bb=28 419 396 775,width=0.5\linewidth]{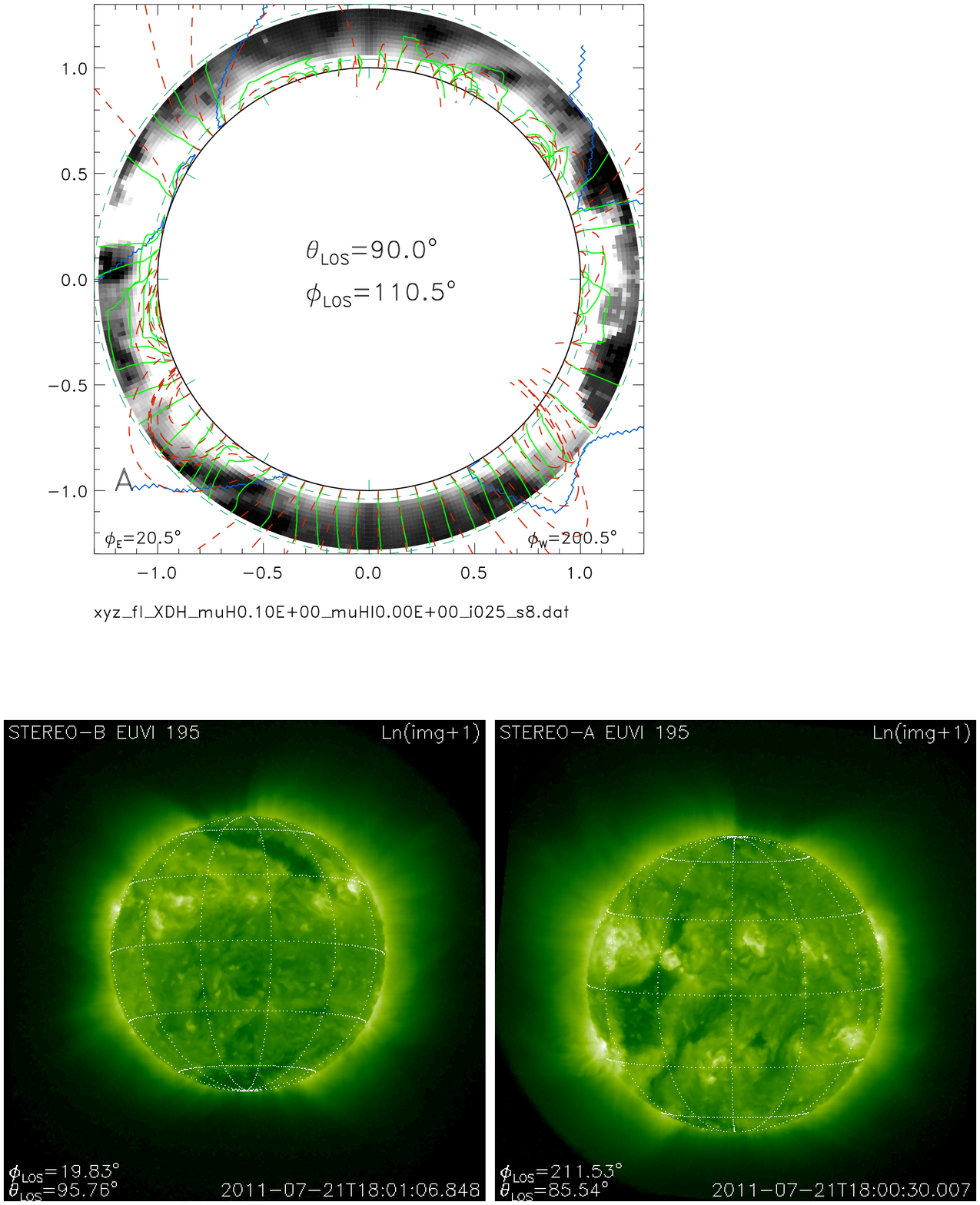}\put(-210,205){\Large{(a)}}
\includegraphics*[bb=28 419 396 775,width=0.5\linewidth]{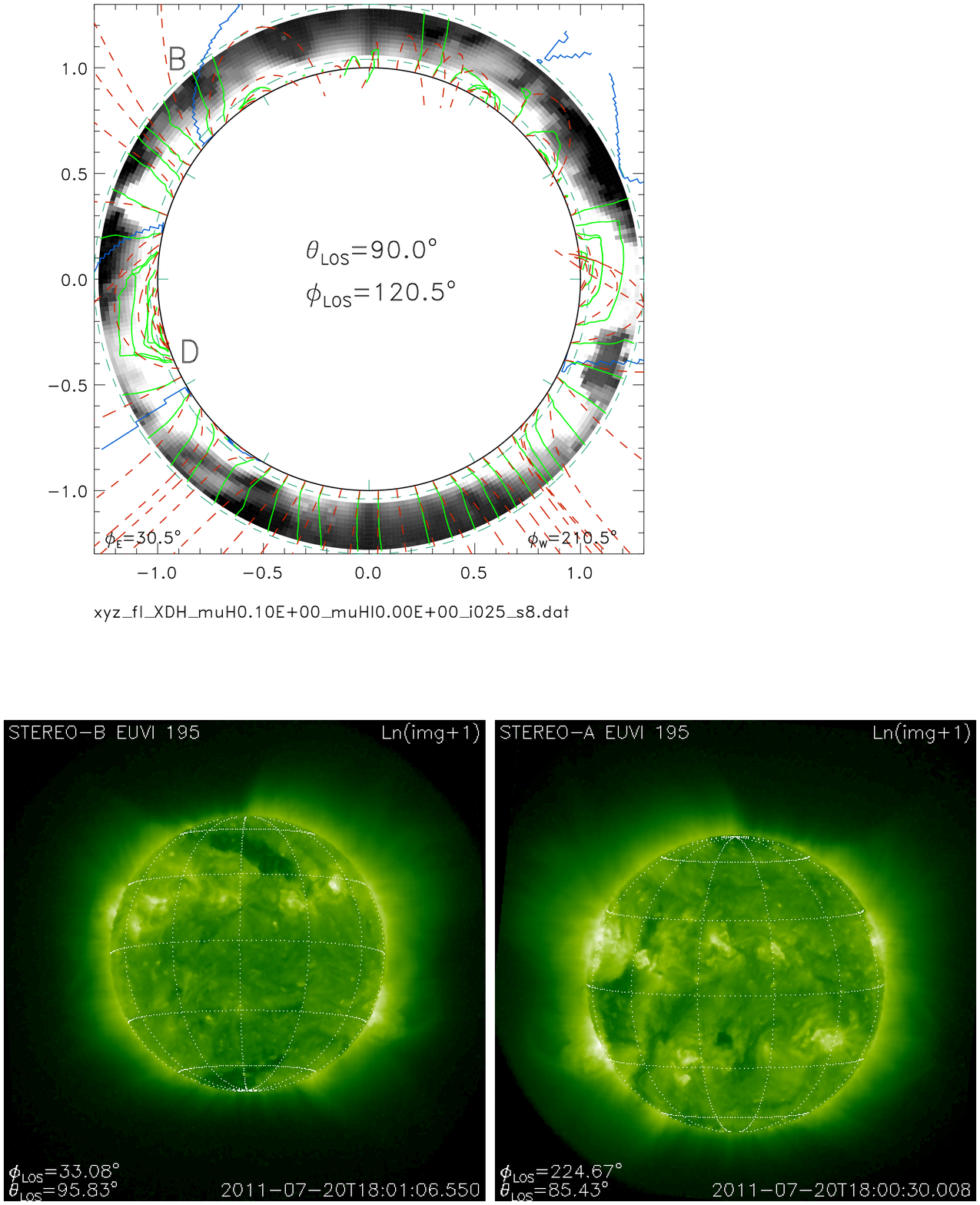}\put(-210,205){\Large{(b)}}\\
\includegraphics*[bb=28 419 396 775,width=0.5\linewidth]{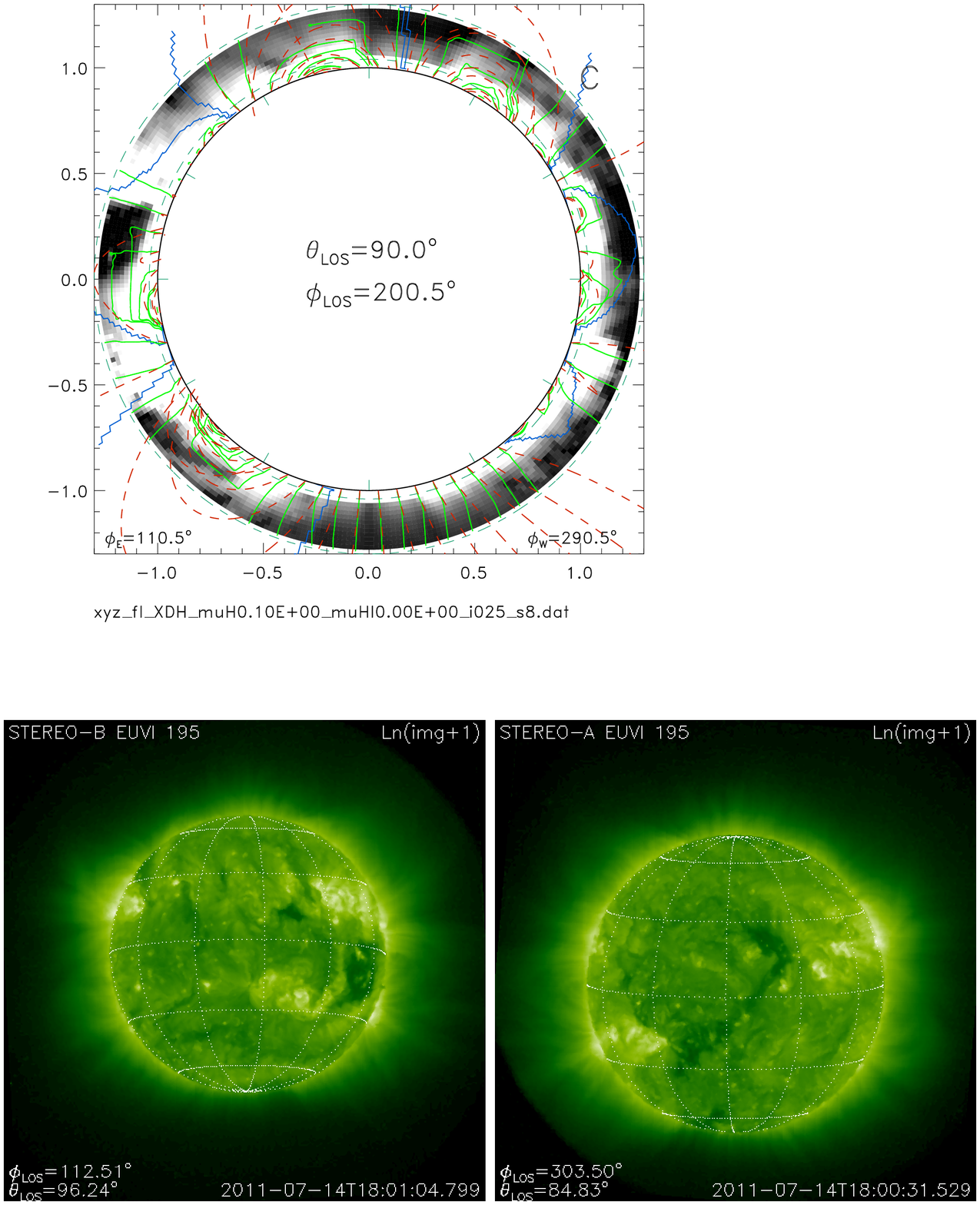}\put(-210,205){\Large{(c)}}
\includegraphics*[bb=88 383 419 703,width=0.5\linewidth]{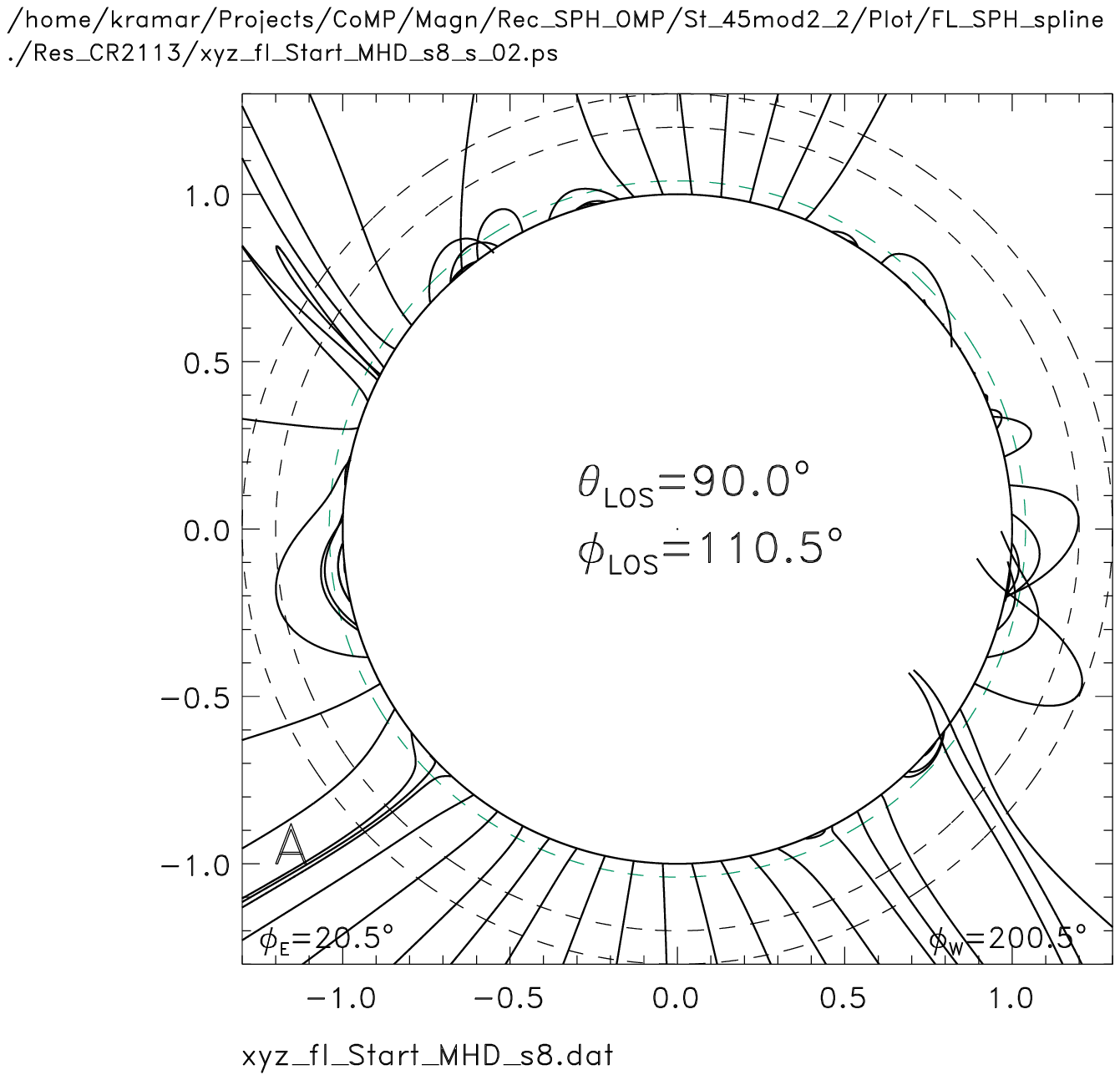}\put(-210,205){\Large{(d)}}
\caption{
Panel (a), (b), and (c) show the magnetic field lines of the coronal fields of CR 2112 derived from tomographic inversion (green lines) and the MHD model (red lines) used as the initial field configuration of the tomographic inversion, originating from Carrington longitudes $\pm90^\circ$ of $110.5$, $120.5$, and $200.5$, respectively. 
The gray filled areas within $[1.06;1.28] R_\odot$ show the reconstructed electron density structure in these cross-sections. Blue lines mark boundaries between regions with open and closed magnetic field lines for the MHD model. The area between the inner and outer dashed circles indicate a FOV region where CoMP input data are used. The animation with a full set of Carrington longitudes is available in the electronic supplemental material accompanying this paper. Panel (d) shows magnetic field lines of the MHD model for CR 2113 originating at Carrington longitude of $20.5^\circ$ and $200.5^\circ$, as in panel (a).} 
\label{Fig_Magn_Rec}
\end{figure}

Figure \ref{Fig_Magn_Rec} shows the LOS projections 
of magnetic filed lines originated at Carrington longitudes $\pm90^\circ$ from LOS directions of $110.5$, $120.5$, and $200.5^\circ$ during CR 2112. In this figure, green lines represent reconstructed field lines derived from the vector tomographic inversion based on CoMP data, and the red dashed lines correspond to a MHD model \citep{Mikic_2007,Lionello_2009}
\footnote{\url{www.predsci.com}} 
\footnote{The MHD model used as our initial field may not be the latest result on the PredSci website.}
that was used as the starting field for the inversion. 
Grayscale areas in the background within $[1.06;1.28] R_\odot$ show the reconstructed electron density. % in the corresponding cross sections. 
Maps of the whole reconstructed 3D fields are shown in the on-line version of this paper. 

The jagged appearances of some of the field lines in the reconstructed model can be attributed to errors in 3D density and temperature reconstructions, the decreasing magnetic field strength as a function of height which makes the regularization less effective at larger heliocentric distances, and reconstruction errors near outer boundary of the reconstruction domain due to limited field of view (FOV, $1.3 R_{\odot}$) of the input polarization data. Nevertheless, the reconstructed fields possess sufficient details to analyze the large scale structure of the coronal magnetic fields. It is interesting to note that the reconstructed fields are in general qualitatively similar to the starting fields except in some regions such as those labelled as A, B, and C, in panel 2(a), 2(b), and 2(c) respectively, where the tomographic inversion resulted in an open field while the starting fields were in a closed configuration.

\begin{figure}[h]
\includegraphics*[bb=67 370 546 698,width=0.49\linewidth]{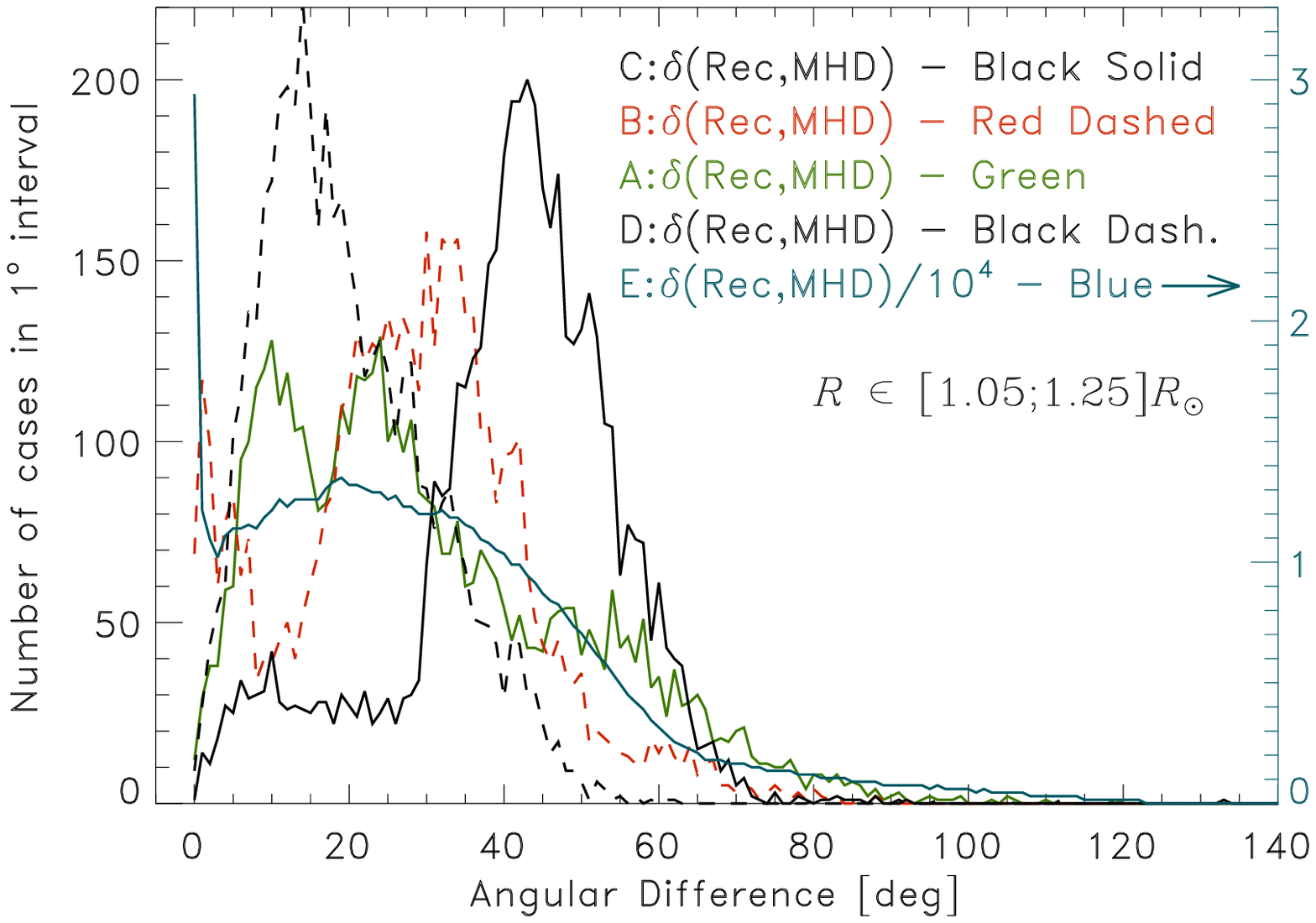}%{DiffA_hist_Reg_r005_130.ps}
\includegraphics*[bb=67 370 546 698,width=0.49\linewidth]{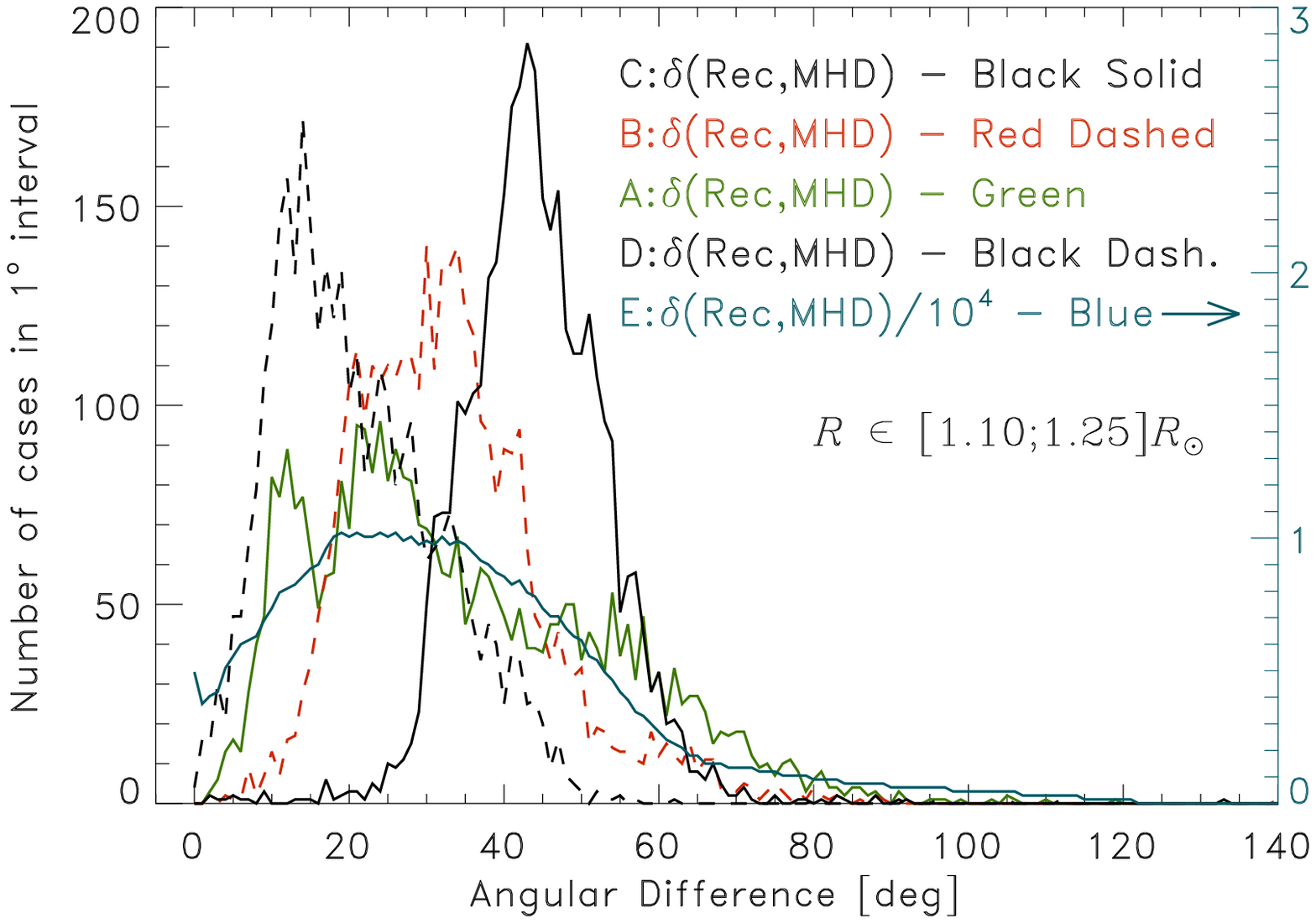}%{DiffA_hist_Reg_r110_130.ps}
\caption{
Distributions of the angular differences between the tomography reconstructed magnetic field vectors and 
that of the MHD model used as the starting field of the tomographic inversion in  regions A (green), 
B (dashed red), C (black), D (black dashed), and E (blue). 
The distributions are within 
$\pm10^\circ$ in longitude and latitude from longitude and latitude of 
$(20.5^\circ;-40.5^\circ)$ (region A), $(30.5^\circ;46.5^\circ)$ (B), $(290.5^\circ;40.5^\circ)$ (C), 
and $(30.5^\circ;-20.5^\circ)$ (D). 
Region E covers entire solid angle of $4\pi$. 
The distributions are within heliocentric distances of $[1.05;1.25]$ (left) and $[1.10;1.25] R_\odot$ (right). } 
\label{Fig_DiffA_Magn_Rec_Hist_Reg}
\end{figure}

%\textcolor{blue}{
While the visual comparison of the magnetic field configurations between 
the MHD and tomographic inversion models in Figure \ref{Fig_Magn_Rec} are instructive, 
a more quantitative presentation of the model differences is shown in Figure \ref{Fig_DiffA_Magn_Rec_Hist_Reg} 
and Table \ref{Table_StDev}. 
In the figure, 
%the \textcolor{blue}{distribution} histograms of the angles 
the histograms of the distributions of the angles, $\delta$,  
between the magnetic field vectors derived from the two methods in volumes 
within $[1.05;1.25]$ and $[1.10;1.25] R_\odot$ in heights, 
and within $\pm 10^{\circ}$ in longitude and latitude around regions A, B, C, and D are plotted in conjunction with that of the entire solid angle volume of $4\pi$ (curve E). 
Region D, as shown in Figure \ref{Fig_Magn_Rec}(b), is a region in which the two models agree fairly well. 
Therefore, its angular difference histogram serves as a measure of the discrepancy between 
the MHD and tomography models. 
We found that the distribution of the angular difference in the reference region D peaks around  $14^{\circ}$. 
On the other hand, the magnetic field orientation difference between the two models are larger in regions A, B, and C, with their histograms 
%\sout{peaking at $25^{\circ}$, $25^{\circ}$, and $40^{\circ}$, respectively} 
having several peaks above $20^\circ$. 
%\textcolor{blue}{
Histogram E have two peaks at about $0$ and $20^\circ$. 
The peak around $0^\circ$ (no change in the field) is predominantly due to contributions from voxels below $1.1R_\odot$. 
This could be because the real field indeed very close to the starting field at lower heights, 
or the tomography method is less sensitive close to the inner boundary of the reconstruction domain 
as observed earlier in scalar field tomography for the coronal electron density \citep{Kramar_2009}. %}
%\textcolor{blue}{
%The relatively smooth peak at about $20^\circ$ demonstrates the presence of regions 
%where the two models agree fairly well and regions where the tomography resulted in different field configuration that the starting field. }
%
%\textcolor{blue}{
Standard deviations, %$\sigma=\sqrt{\sum\limits_{i=1}^N{\delta_i^2}/N}$, 
$\sigma=\sqrt{\sum_{i=1}^N{\delta_i^2}/N}$, of $\delta$ for the regions of interest 
are summarized in Table \ref{Table_StDev}. 
Region D has the smallest value of $\sigma$ demonstrating fairly well local agreement between the two models. 
Region C may contain small streamer or pseudo-streamer 
\citep{Zhao_Webb_2003JGRA,Wang_YM_2007ApJ,Wang_YM_2015ApJ,Abbo_2015SoPh} 
where the sign of the field radial component, $B_r$, changes abruptly in the angular dimensions %with $\delta>90^\circ$ 
causing the large value of $\sigma$. 
Field differences in region A are predominantly confined within narrow longitudinal range 
compared to the region width. 
Indeed, cross-sections for $10.5$ and $30.5^\circ$ exhibit a fairly well local agreement 
between the two models near region A. 
This explains a comparable values of $\sigma$ for regions A and E. 
Region B exhibits the same sign of $B_r$ causing lower value of $\sigma$. %}

%A larger value of $\sigma$ for region E, compared with those for A and B, 
%is due to small amount of voxels with large values of $\delta$ and specifics of regions A and B. 
%Field differences in region A are predominantly confined within narrow longitudinal range centered at $\sim20^\circ$ (see Figure \ref{Fig_Magn_Rec}). 
%The field in region B exhibits the same polarity for both models causing $\delta$ to be less than $90^\circ$. 
%Region C may contains small streamer or pseudo-streamer where the field polarity changes gradually leading to $\delta>90^\circ$. 
%Therefore, a small change in the field topology leads to larger value of $\sigma$ in region C than for A and B. }

\begin{figure}[h]

\includegraphics*[bb=64 296 550 620,width=0.49\linewidth]{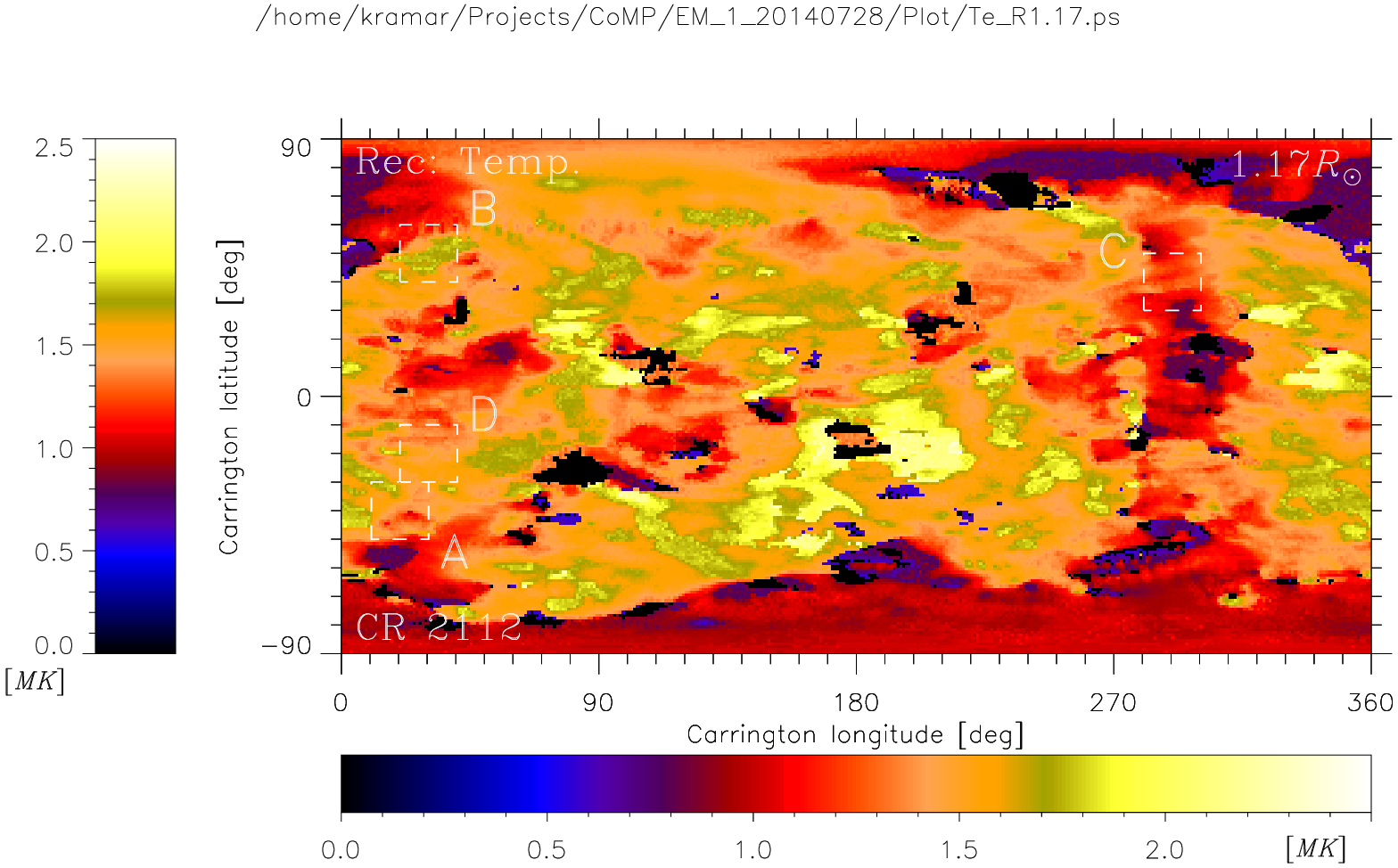}\\
\includegraphics*[bb=64 296 550 620,width=0.49\linewidth]{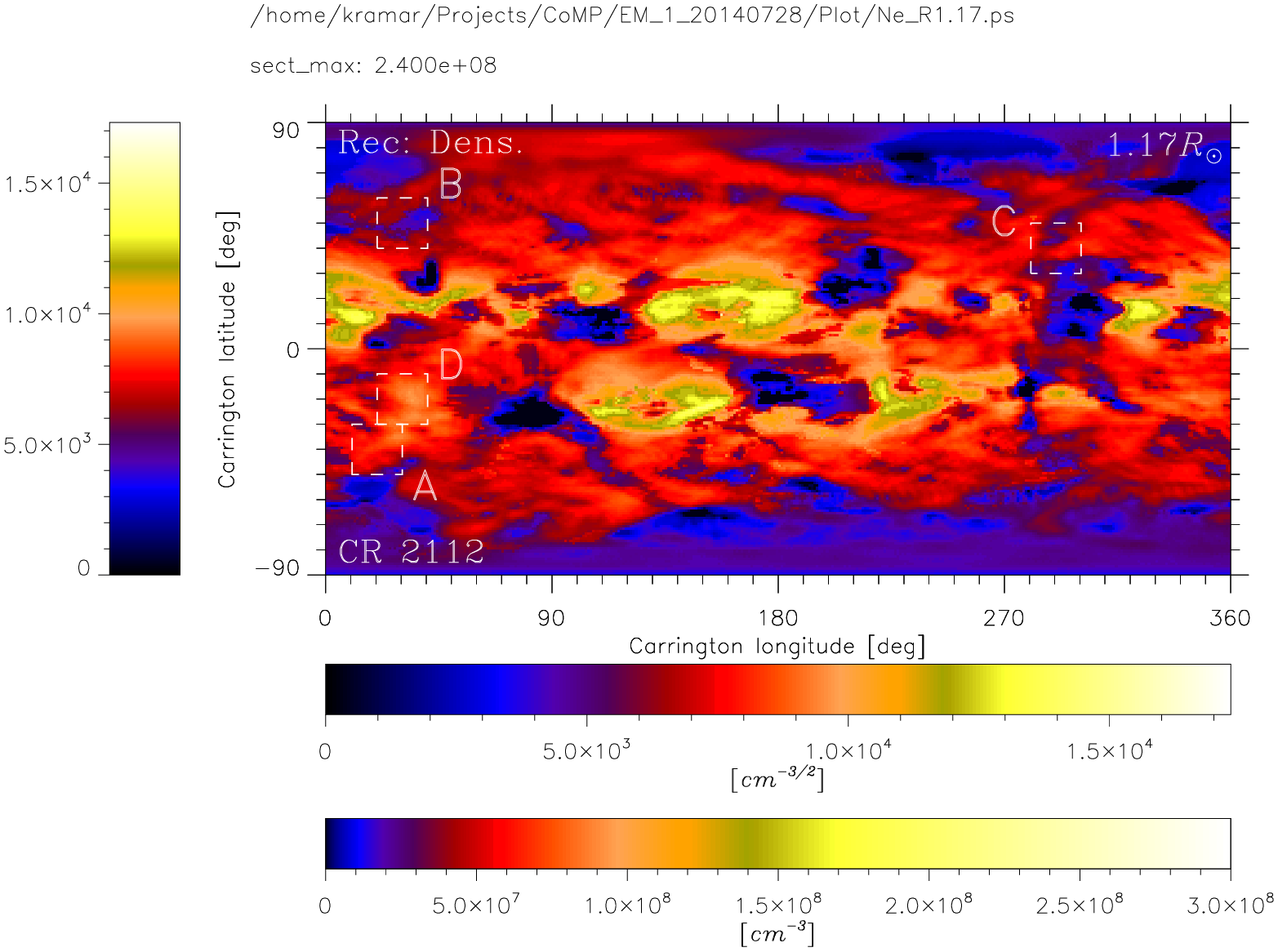}\\
\includegraphics*[bb=64 296 550 620,width=0.49\linewidth]{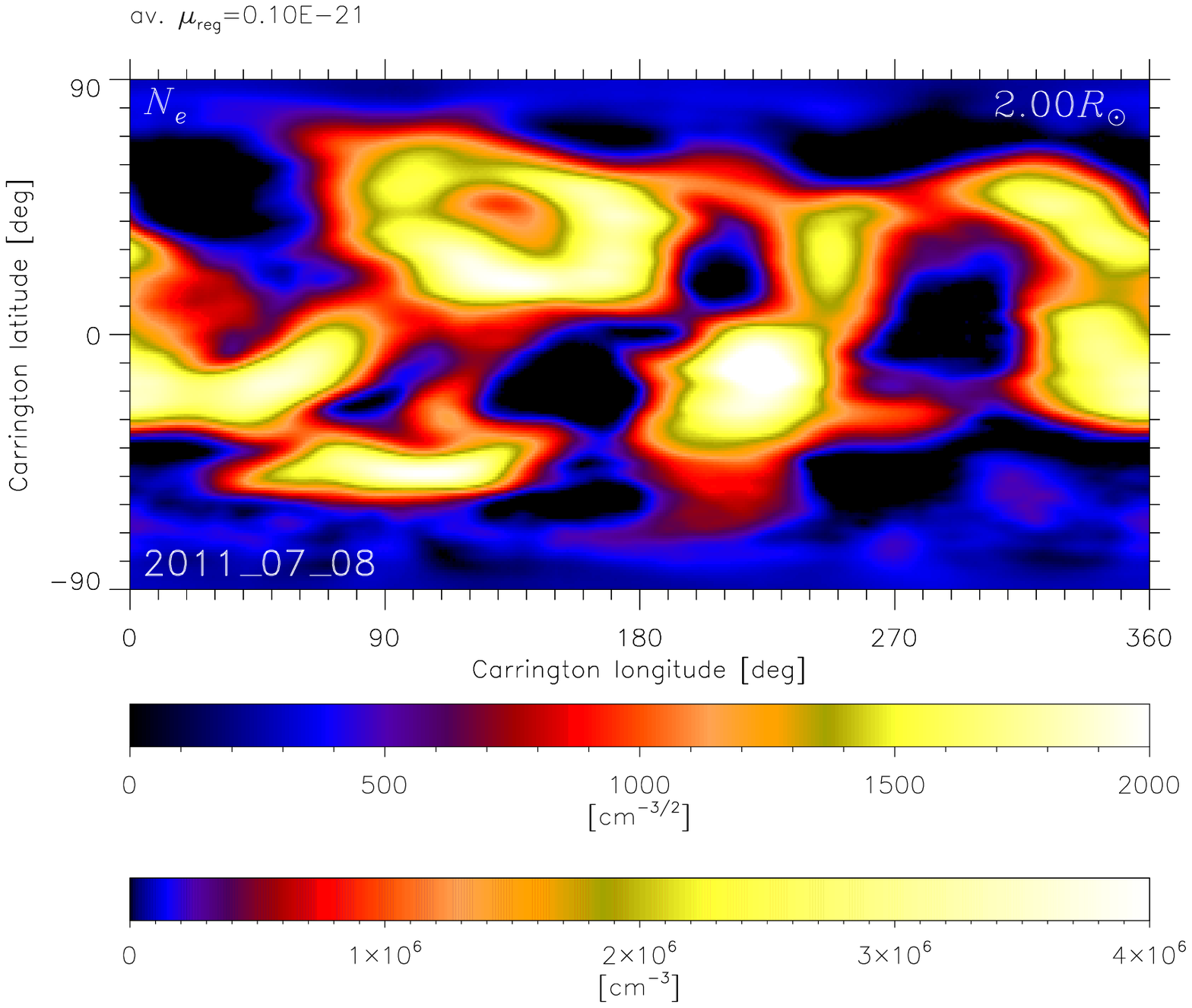}
\caption{The spherical cross-section of the reconstructed 3D temperature (top) and electron density (middle) at heliocentric distance of $1.17 R_\odot$ obtained with EUVI Tomography, and electron density map at $2.0 R_{\odot}$ obtained with STEREO/COR1 tomography (bottom).}
\label{Fig_NeTe_Rec}
\end{figure}

Although we have used the MHD simulated model as our reference model for evaluation of the efficacy of the tomographic inversion method, at this point it should be stressed that it is currently impossible to perform independent verification of the tomographic inversion results --- MHD models based on observed photospheric magnetic fields inputs were derived with a completely different set of assumptions, and are subjects to its own limitations and uncertainties, and at this point which method provides more accurate result cannot be assessed. However, there are other coronal observations and reconstructions that in certain situation can help us assess the validity of the tomographic inversion results. For examples, coronal holes should be associated with open field regions, and current sheet structures are known to be associated with closed dipole-like magnetic field structures \citep{Kramar_2014}. Therefore we have examined the 3D electron density structures of CR 2112 at $1.17R_{\odot}$ and $2.0R_{\odot}$, as shown in Figures \ref{Fig_NeTe_Rec}(b), \ref{Fig_Magn_Rec}, 
and \ref{Fig_NeTe_Rec}(c), to see if they can provide additional information to help us assess the validity of the tomographic inversion results. 

For region A, the density map at $2 R_{\odot}$ showed a small low density region, inconsistent with the existence of a current sheet with a underlying closed field region. Thus, the open field configuration of the tomographic inversion still cannot be ruled out. 
Also, it should be noted that the photospheric magnetic fields used as the boundary condition of the MHD model contain data that were collected about 3/4 solar rotation before the reference date of the region A of CoMP observation, and may represent an older field configuration. Examination of the MHD model for the next CR 2113 as shown in Figure \ref{Fig_Magn_Rec}(d) reveals that it has a magnetic field configuration similar to that of the tomography inverted results in region A. 
%Thus, the open field configuration derived from tomographic inversion in region A is not in complete disagreement with MHD simulated results. 

\begin{figure}[h]
\includegraphics*[bb=2 40 298 335,width=0.5\linewidth]{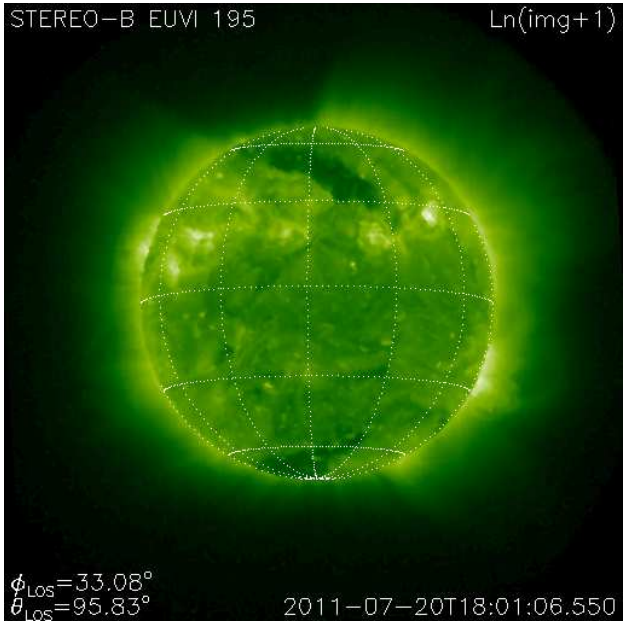}
\caption{
STEREO-B/EUVI image taken on July 20, 2011 when CoMP POS was at the longitude of $33^\circ$. 
Central meridian of the EUVI image is located at $33^\circ$. 
White dots show carrington coordinate grid for every $30^\circ$.}
\label{Fig_EUVI_LOS30deg}
\end{figure}

For region B, the density map of this region at $\sim1.2 R_{\odot}$ shows a small low density region nearby, while the density map at $2.0 R_{\odot}$ 
shows an extended low density region (Figure \ref{Fig_NeTe_Rec}), 
strongly suggesting the presence of a coronal hole at this location. Additional support for this interpretation can be found in the EUVI 195 \AA \ images of STEREO B. During July 2011, the angle between STEREO B and the Earth was approximately $90^\circ$. Therefore, the east limb in CoMP observations are observed simultaneously in the central meridian in the STEREO EUVI images. Figure \ref{Fig_EUVI_LOS30deg} shows EUVI-B 195 \AA \ image with a central meridian located at $\phi_{LOS} = 33^\circ$. It shows a coronal hole at longitude of $30^\circ$ that spread up to latitude of at least $50^\circ$, providing further support for the open field configuration derived from the tomographic inversion.

For region C, the density cross-section in Figure \ref{Fig_NeTe_Rec}(b) shows a narrow structure 
running along the east-west direction. 
This type of structure may be associated with small-scale streamers or pseudo-streamers 
\citep{Zhao_Webb_2003JGRA,Wang_YM_2007ApJ,Wang_YM_2015ApJ,Abbo_2015SoPh} 
with a small closed field region in the core. 
While the size of these small scale streamer cores may be comparable to 
the resolution of our tomographic reconstruction, 
Figure \ref{Fig_Magn_Rec}(c) indeed shows a small closed loop near the limb embedded inside an open field structure, 
suggestive of the existence of a small scale streamer or pseudo-streamer. 

%For region C, \sout{we do not feel that the density maps can be used to infer the magnetic field configuration in this region with any certainty} \textcolor{blue}{
%the density cross-sections on Figures \ref{Fig_NeTe_Rec}(b) and \ref{Fig_Magn_Rec}(c) exhibit 
%a narrow in latitudinal direction density structure. 
%Such type of structures might be associated with small streamer or pseudo-streamer 
%\citep{Zhao_Webb_2003JGRA,Wang_Sheeley_1992,Abbo_2015SoPh} 
%with very small closed field region in the core which angular size is compared with reconstruction resolution.} 

\section{Discussions and Conclusions}

We have performed the first ``direct observation'' of the 3D coronal magnetic field using a combination of vector tomographic inversion of IR CEL {\it linear} polarization data and scalar tomographic inversion of CEL intensity data in the UV wavelengths. We compared the tomography inverted coronal magnetic fields with those constructed by MHD simulation based on observed photospheric magnetic fields of CR 2112 and 2113 as well as with the STEREO/EUVI 195 \AA \ image and with the global 3D coronal electron density structure obtained by tomography based on STEREO/COR1 data. This paper present a first look and preliminary analysis of the results. A comprehensive analysis of the 3D coronal fields and its relationship to the temperature and density structures, as well as comparison with other coronal models will be presented in a future paper. 

Our analysis so far suggest that tomographic inversion based on CEL polarization data and MHD simulations based on photospheric boundary conditions are both capable of deriving the large-scale coronal magnetic field structures, albeit they are subject to different constraints and uncertainties and possess different type of errors. For examples, global extrapolation methods such as global MHD, potential field (PFSS), and force-free field extrapolations use {\it photospheric} field data collected {\it over a solar rotation} period as a boundary condition (see Section 1 in \cite{Kramar_2014} and references therein) while the vector tomography primarily uses direct {\it coronal} data that requires observations from {\it only half of a solar rotation} period. 
Furthermore, in the tomography, the coronal fields are retrieved without any presumption (such as the potential or force-free conditions) on the conditions of the fields. 
The tomography requires simultaneous observations of the same object through multiple sight lines. 
Nevertheless, although current instrumentation capabilities do not allow for simultaneous observation of the CEL polarization from more than one sight line, 
%(this statement can also be applied to the photospheric boundary field data needed for coronal MHD and PFSS models), 
data obtained over an extended period of time as the Sun rotates can be interpreted as data from multi-sight-line observations for slowly evolving  coronal magnetic field structures. Therefore, the coronal magnetic field structures derived from the tomographic inversion method should be a fair representation of slowly evolving, large scale magnetic structures. 

%\textcolor{blue}{
Depending on the positions of a coronal structure relative to an observer, 
many regions close to equator are visible during a period less than half a solar rotation. 
Therefore, although the tomography method is limited in reconstructing a Coronal Mass Ejection (CME) structure, 
the method can be used for reconstruction of pre- and post-CME global 3D coronal structures \citep{Kramar_2011}. 
This will help us to understand properties of CMEs and their ultimate origin and precursors. %}

This research utilizes polarization data from only one spectral line. To further improve the accuracy and reliability of this technique, we plan to incorporate additional polarization data from the Fe XIV 5303 \AA\ line from the visible CoMP instrument at Lomnicky Stit Observatory in Slovakia \citep{Rybak_2010nspm}. We also plan to include STEREO/EUVI on-disk data to improve the accuracy of the temperature and density inversion. Finally, the current vector tomographic inversion utilizes only CEL linear polarization data which only yield information about the direction of the magnetic field projected in the POS. As the sensitivity of the instruments continue to improve, incorporation of the CEL line intensity and circular polarization data that carry information about the height of the data sources, and the strength of the longitudinal component of the magnetic fields will further improve the accuracy and sensitivity of the vector tomographic inversion technique. Future projects, such as the Daniel K. Inouye Solar Telescope currently under construction on the summit of Haleakala \citep{Elmore_2014SPIE}, or the Coronal Solar Magnetism Observatory (COSMO) proposed by the High Altitude Observatory and its partners, will be able to provide the full-Stokes data for the full-vector tomographic inversion, leading to a more accurate 3D reconstruction of the coronal magnetic field. We envision that with improved observing capabilities in the future, in particular, with the implementation of multiple sight line observing capabilities similar to the STEREO mission, tomographic inversion method should provide the most reliable coronal magnetic field data to validate other coronal field inference methods and theoretical models.

\section*{Acknowledgment}
The authors would like to thank the anonymous referee and scientific editor, Dr. Rekha Jain, 
for helpful suggestions for the presentation of the data, and careful reading of the manuscript that greatly improve the quality of this paper, as well as Phill Judge and Roberto Casini of the High Altitude Observatory for providing the CLE codes used to compute the emission coefficients of the Stokes I,Q,U and V parameters for the forbidden coronal emission lines. 
MK thanks Bernd Inhester for very helpful advises leaded to this paper and Peter Young for helpful information about CHIANTI database. 
The National Center for Atmospheric Research is sponsored by the National Science Foundation (NSF). 
This research was supported by NSF National Space Weather Program grant number AGS0819971.

%\bibliographystyle{apj}

%\tracingmacros=2
%\bibliography{biblio}

\clearpage

%------------------
% Figure 1

\clearpage

%------------------
% Figure 2

\onecolumn

%------------------
% Figure 3

%The blue solid line shows the angular difference distribution of the entire reconstruction domain within $1.25R_\odot$ sphere.

%------------------
% Figure 4
%
%\begin{figure}
%\includegraphics*[bb=63 299 555 621, width=0.5\linewidth]{f4.ps}
%{/home/kramar/Projects/MyPubl/Paper_CoMP/lehua/2011_07_08/01u_06_SPH_Rec_gw/SphSect_rec_R2.00_mu0.10E-21_sqrt2.ps}
%\caption{Spherical cross-section of the reconstructed electron density in square-root scale at heliocentric distance of $2 \ \mathrm{R}_\odot$. 
%The reconstruction is obtained by tomography based on COR-1 data obtained during July, 2011 (CR 2112).}
%\label{Fig_COR1_SphSect}
%\end{figure}

%------------------
% Figure 4

%\begin{figure}[h]
%\includegraphics*[bb=67 370 547 698,width=0.49\linewidth]{./Plot_Diff/DiffA_hist.ps}
%\caption{
%Distributions of angular differences over all reconstruction domain 
%between reconstructed and starting magnetic field vectors (black), 
%between the PFSS and the starting magnetic field vectors (red), 
%and between the starting field and PFSS (green). }
%\label{Fig_DiffA_Magn_Rec_Hist}
%\end{figure}

\clearpage

\begin{table}[]
\centering
\begin{tabular}{|l|l|l|l|l|l|l|}
\hline
\multicolumn{2}{|l|}{Region}                       & A    & B    & C    & D    & E    \\ \hhline{|=|=|=|=|=|=|=|}
\multirow{2}{*}{$[1.05;1.25]R_\odot$} & $\sigma$   & 36.4 & 32.0 & 43.5 & 23.2 & 38.3 \\ \cline{2-7}
                                      & $\delta_m$ & 24.0 & 30.0 & 43.0 & 14.0 & 0.0  \\ \hhline{|=|=|=|=|=|=|=|}
\multirow{2}{*}{$[1.10;1.25]R_\odot$} & $\sigma$   & 39.4 & 35.7 & 45.4 & 24.2 & 41.4 \\ \cline{2-7}
                                      & $\delta_m$ & 24.0 & 30.0 & 43.0 & 14.0 & 19.0  \\ \hline
\end{tabular}
\caption{Standard deviation, $\sigma$, and position of the histogram maximum value, $\delta_m$, 
for the analyzed regions.}
\label{Table_StDev}
\end{table}

\end{document}